\documentclass{INTERSPEECH2023}
\usepackage{graphicx}
\usepackage{amsmath}
\usepackage{amsfonts,amssymb}
\usepackage{float}
\usepackage{array}
\usepackage{booktabs}
\usepackage{epstopdf}
\usepackage{color}

\usepackage{stfloats}
\usepackage{hyperref}

\interspeechcameraready

\name{Kun Song$^{1,\dag}$\thanks{$^{\dag}$Work done during an internship at ByteDance. Lei Xie is the corresponding author.}, Yi Ren$^2$, Yi Lei$^1$, Chunfeng Wang$^2$, Kun Wei$^1$, Lei Xie$^1$, Xiang Yin$^2$, Zejun Ma$^2$}
\title{StyleS2ST: Zero-shot Style Transfer for Direct Speech-to-speech Translation}
\vspace{-6pt}
\address{
  $^1$Audio, Speech and Language Processing Group (ASLP@NPU)\\School of Computer Science,
  Northwestern Polytechnical University, Xi’an, China\\
  $^2$ByteDance}
  \vspace{-6pt}
\email{kunsong.npu.se@gmail.com, lxie@nwpu.edu.cn}
\vspace{-6pt}
\begin{document}
\vspace{-6pt}
\maketitle
\vspace{-6pt}
\begin{abstract}
\vspace{-3pt}
Direct speech-to-speech translation (S2ST) has gradually become popular as it has many advantages compared with cascade S2ST. However, current research mainly focuses on the accuracy of semantic translation and ignores the speech style transfer from a source language to a target language. The lack of high-fidelity expressive parallel data makes such style transfer challenging, especially in more practical zero-shot scenarios. To solve this problem, we first build a parallel corpus using a multi-lingual multi-speaker text-to-speech synthesis (TTS) system and then propose the \textit{StyleS2ST} model with cross-lingual speech style transfer ability based on a style adaptor on a direct S2ST system framework. Enabling continuous style space modeling of an acoustic model through parallel corpus training and non-parallel TTS data augmentation, StyleS2ST captures cross-lingual acoustic feature mapping from the source to the target language. Experiments show that StyleS2ST achieves good style similarity and naturalness in both in-set and out-of-set zero-shot scenarios.

\end{abstract}
\noindent\textbf{Index Terms}:  direct speech-to-speech translation, style transfer, zero-shot

\vspace{-6pt}

\section{Introduction}
\vspace{-3pt}
In recent years, direct speech-to-speech translation (S2ST) has attracted more and more attention~\cite{DBLP:conf/interspeech/JiaWBMJCW19, DBLP:conf/icml/JiaRRP22, DBLP:conf/acl/LeeCWGPMPAHTPH22}. Conventional S2ST~\cite{DBLP:journals/taslp/NakamuraMNKKJZYSY06} is usually based on cascade systems, including automatic speech recognition (ASR), machine translation (MT), and text-to-speech synthesis (TTS) modules, while direct S2ST aims to integrate the above modules into a unified model for directly synthesizing target language speech translated from the source language. Compared with the cascade system, direct S2ST takes advantage of a more elegant pipeline, which avoids relying on intermediate text and obtains less error propagation. Recent direct S2ST works have achieved promising performance by introducing speech-to-unit translation (S2UT) models~\cite{DBLP:conf/acl/LeeCWGPMPAHTPH22, DBLP:conf/interspeech/PopuriCWPAGHL22}, which leverage discrete self-supervised units, such as features from HuBERT~\cite{DBLP:journals/taslp/HsuBTLSM21}, as an intermediate representation bridging different languages in S2ST. These approaches adopt self-supervised pre-trained modules and data augmentation strategies to improve semantic accuracy, leading to state-of-the-art (SOTA) performance. 

Since the unit of self-supervised learning can well represent the semantic information in speech, it helps the model better learn the alignment relationship between source and target languages. Therefore, direct S2ST has achieved semantic translation performance close to or even better than conventional cascade systems. 
However, current S2ST studies mainly focus on linguistic information, while ignoring \textit{speaking styles} (e.g., speaker timbre, intonation, rhythm, and intensity) that accompany the source language speech. The stylistic appearance of speech is complex and contains rich information such as speaker identity and attitude, which significantly affects the expressiveness and naturalness of speech. With the increase of application scenarios for S2ST, it is often critical to transfer the source speaking style to the target language speech in many scenarios, such as movie dubbing translation. While S2ST with style transfer is challenging due to the lack of paired multi-lingual speech data from the same speaker. Moreover, in real-world applications, it is an absolute limitation if only transferring the style seen during training since there may exist diverse style expressions in source language speech. Therefore, the generalization ability is essential in S2ST for an arbitrary unseen style from a new speaker, which we call the \textit{zero-shot} style transfer of S2ST in this paper. 

In this work, firstly, we would like to define the basic specifications of \textit{style} in S2ST scenarios. Specifically, style here includes low-level characteristics such as speaker timbre, pitch, intensity, and rhythm, while high-level characteristics are always corresponding to speaker identity. We consider different speakers usually have distinguished speaking styles. Given the above assumption, this paper mainly focuses on zero-shot style transfer for S2ST, which has several challenges as follows:


\vspace{-2pt}
\begin{itemize}
\item The stylistic expression and semantic content should be \textit{disentangled} for style transfer in S2ST, while current works lack such ability since they only concern with the semantic translation. Moreover, style and linguistic contents are heavily entangled in speech and they are not easily to be disentangled. 

\item To conduct style transfer in direct S2ST, high-fidelity expressive paired data of source and target language speech from the same speaker is necessary but extremely hard to obtain.

\item In zero-shot scenarios, it usually requires enormous amounts of speech data of diverse speaking styles to build a complete continuous speaker space. However, the demand for paired speech heavily increases the difficulty of zero-shot style transfer which desires good out-of-set speaker generalization.
\vspace{-1pt}
\end{itemize}

\vspace{-2pt}
To construct parallel bilingual speech data, an effective solution is to leverage a TTS model for pseudo-data pairs augmentation~\cite{DBLP:conf/interspeech/JiaWBMJCW19, DBLP:conf/icml/JiaRRP22}. Previous works usually synthesize monolingual speech from text extracted by speech-to-text translation, due to the high cost and limited ability of the pre-trained TTS models. In that way, the constructed pseudo-data pairs realize semantic parallelization, but not the speaker parallel parallelization. Existing S2UT-based S2ST works~\cite{DBLP:conf/acl/LeeCWGPMPAHTPH22, DBLP:conf/interspeech/PopuriCWPAGHL22} mainly use single-speaker unit decoder to achieve single-speaker speech-to-speech translation. Translatotron~\cite{DBLP:conf/interspeech/JiaWBMJCW19}, proposed for S2ST without intermediate representations, tries to achieve style transfer by a speaker encoder, but the performance is far from good enough for applications. Moreover, current direct S2ST systems can not reserve style from source to target speech, while zero-shot style transfer has even not been considered yet.

\begin{figure*}[ht]

	\centering
	\includegraphics[width=16.8cm,height=6.3cm]{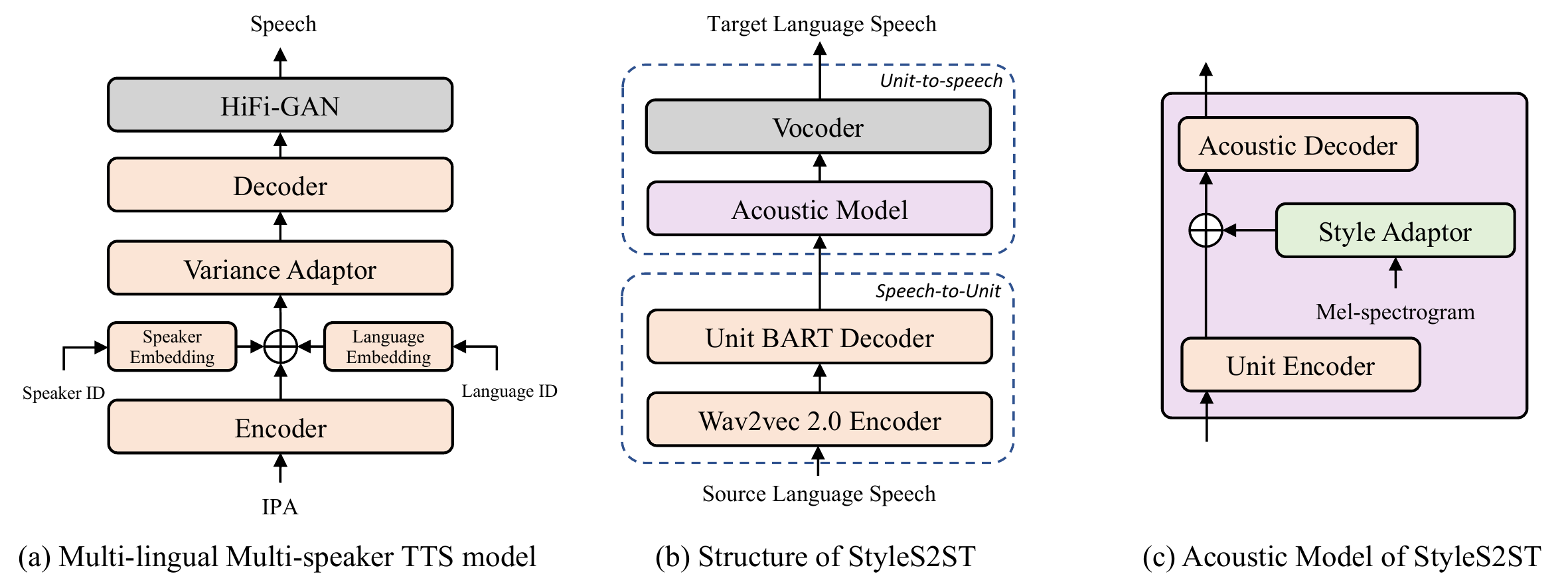}
	\vspace{-6pt}
        \caption{Architecture of our proposed methods, consists of a TTS model for building the bilingual parallel corpus and our proposed S2ST system -- StyleS2ST.}
        \vspace{-8pt}
	\label{model_dspgan}
\end{figure*}

In this paper, we propose to conduct zero-shot style transfer for direct speech-to-speech translation. To construct a high-quality bilingual parallel corpus, we leverage a pre-trained multi-lingual multi-speaker TTS system taking the bilingual parallel texts as input. With the synthetic speech parallel in both semantics and speaking style, we build the proposed method based on a SOTA direct S2UT English-to-Chinese (En2Zh) system~\cite{DBLP:conf/interspeech/PopuriCWPAGHL22}. To achieve zero-shot style transfer, we adopt a pre-trained \textit{style adaptor} based on a speaker embedding model in our S2UT system. The contributions of this paper can be summarized as follows. 1) To the best of our knowledge, this is the first attempt for zero-shot style transfer in direct S2ST. 2) We leverage a multi-speaker multi-style TTS system to construct parallel bilingual speech of the same semantic content and speaking style. 3) We propose to utilize a style adaptor to capture the speaking style of arbitrary unseen speakers for achieving zero-shot style transfer in direct S2ST. Experimental results show that the proposed approach, dubbed \textit{StyleS2ST},  can successfully conduct zero-shot end-to-end speech-to-speech translation with the preservation of speaking style from the source to the target languages. Our audio samples are available on the demo website.\footnote{
\url{https://StyleS2ST.github.io/StyleS2ST}}
\vspace{-6pt}
\section{Methods} 
\vspace{-3pt}
Our goal is to realize zero-shot style transfer in a direct S2ST system. In this section, we will introduce the proposed pseudo-data pairs construction process and the structure of our direct S2ST system based on S2UT. Next, we present our proposed style adaptor and the enhanced acoustic decoder to achieve style transfer and zero-shot, respectively.

\vspace{-6pt}
\subsection{Corpus Construction}
\vspace{-3pt}
To train the StyleS2ST, we construct a speaker-parallel En2Zh corpus through a multi-lingual multi-speaker TTS system. Figure~\ref{model_dspgan}(a) shows our TTS structure based on the acoustic model of Fastspeech 2~\cite{DBLP:conf/iclr/0006H0QZZL21} and the vocoder of HiFi-GAN~\cite{DBLP:conf/nips/KongKB20}. Firstly, we adopt the international phonetic alphabet (IPA), widely applied in cross-lingual TTS and voice conversion (VC)~\cite{DBLP:conf/interspeech/ZhangWZWCSJRR19, DBLP:journals/corr/abs-2011-06392}, as input to model the shared speech space of different languages. Then, we use an encoder to encode the input phoneme sequence into a hidden representation, which is concatenated with injected speaker embedding and language embedding to synthesize speech in the target language and desired speaking style. Due to the lack of parallel corpus, each speaker has only English or Chinese data. Still, using IPA as input allows us to obtain bilingual high-fidelity synthesized speech for different speakers. To evaluate and filter the synthesized speech for training S2ST, we utilize word error rate (WER) and perceptual evaluation of speech quality (PESQ)~\cite{DBLP:conf/icassp/RixBHH01} to check the semantic errors and speech quality. Any generated speech with a performance below the threshold is discarded from the corpus.

\vspace{-6pt}
\subsection{Structure of StyleS2ST System}
\vspace{-3pt}
Following the setup of~\cite{DBLP:conf/icml/JiaRRP22}, we extract the hidden representations of the pseudo-data pairs through the pre-trained HuBERT model, which is trained on a large amount of unlabeled speech data. The extracted hidden representations of speech are clustered by the k-means method to encode speech into discrete units as intermediate representations for the S2UT system. The direct StyleS2ST model composes of speech-to-unit and unit-to-speech modules. In addition, we keep the consecutive repeating units to preserve prosodic information in duration as much as possible. 

As illustrated in Figure~\ref{model_dspgan}(b), the speech-to-unit module employs a pre-trained wav2vec 2.0~\cite{DBLP:conf/nips/BaevskiZMA20} speech encoder and a pre-trained unit BART~\cite{DBLP:conf/acl/LewisLGGMLSZ20} decoder based on the Transformer~\cite{DBLP:conf/nips/VaswaniSPUJGKP17} architecture, following the approach presented in~\cite{DBLP:conf/interspeech/PopuriCWPAGHL22}. In contrast, the unit-to-speech module differs in that we do not directly use the HiFi-GAN vocoder. Instead, we divide the vocoder into two parts: the acoustic model and the vocoder itself. This pipeline is more common in text-to-speech (TTS) and facilitates easier style control. The Fastspeech~\cite{DBLP:conf/nips/RenRTQZZL19} architecture is utilized as the acoustic model, while the DSPGAN~\cite{DBLP:journals/corr/abs-2211-01087} vocoder is used due to its superior robustness in universal scenarios. This is particularly significant for speech quality in zero-shot scenarios.


\vspace{-6pt}
\subsection{Acoustic Model}
\vspace{-3pt}
\subsubsection{Style Adaptor}
\vspace{-3pt}
Retaining the duration features in the speech-to-unit module enables the inputs of the acoustic model to express local style features such as prosody.  To capture global style features, including speaker timbre, we adopt an utterance-level global style adaptor.   However, constructing the speaker space and modeling speaker features for unseen speakers is a critical challenge when implementing zero-shot capability.   A common method is to employ a reference encoder, such as the global style token (GST)~\cite{DBLP:conf/icml/WangSZRBSXJRS18} or variational autoencoder (VAE)~\cite{DBLP:conf/iclr/HsuZWZWWCJCSNP19}, to generate a global style embedding from a reference speech.   Nonetheless, the speaker space requires a significant amount of training data, and acquiring sufficient paired data is challenging.   To address this issue, we utilize a pre-trained style adaptor based on the style embedding extracted from speaker embedding models.   Specifically, we employ the ECAPA-TDNN~\cite{DBLP:conf/interspeech/DesplanquesTD20} speaker embedding model as the style adaptor and pre-train it using a large amount of Chinese and English data.   By training in two languages, we can avoid mismatch issues in speaker space between different languages, which is crucial for our S2UT system.


Since units are frame-level, there is no need for a duration predictor. Instead, we use the units as input to the unit encoder to capture contextual information. Figure~\ref{model_dspgan}(c) illustrates the implementation of the style adaptor module, which extracts a style embedding as an additional condition for the acoustic decoder. This enables style control ability and continuous acoustic space modeling.

%
\vspace{-4pt}
\subsubsection{Enhanced by Unpaired Data}
\vspace{-3pt}
Given the limited number of speakers in the paired data, we employ a large number of unpaired Chinese data to pre-train the acoustic model. Fine-tuning during speaker adaptation can lead to catastrophic forgetting problems~\cite{DBLP:journals/corr/abs-2103-14512}. Therefore, we choose not to fine-tune the pre-trained model on our parallel corpus. Instead, we train the pre-trained model using both Chinese speeches in the parallel corpus and unpaired Chinese data, thereby increasing the amount of training data available.

\vspace{-6pt}
\section{Experiments}
\vspace{-3pt}
\subsection{Datasets}
\vspace{-3pt}
For all the data used in this section, we downsample the audio to 16 kHz. The frame-level features utilize a 20 ms frame shift and a 50 ms frame length.
\vspace{-5pt}
\subsubsection{En2Zh Parallel Corpus}
\vspace{-3pt}
CoVoST 2~\cite{DBLP:conf/interspeech/WangWGP21} is a commonly used speech-to-text translation dataset that contains 512K English speech and corresponding Chinese text data pairs. However, as the English speech data in CoVoST 2 is not high-fidelity and may result in a loss of speech quality when using speaker adaptation in TTS to construct corresponding Chinese speech, we use only the text data in our experiments.   We train a multi-lingual multi-speaker TTS system using an internal dataset consisting of 20 speakers.  Each speaker provides approximately 5,000 utterances and 5 hours of high-fidelity expressive Chinese or English data, with varying styles, including stories, broadcasts, etc., and covering speakers of different genders and ages.   Using the TTS system, we distribute the Chinese texts in CoVoST 2 and corresponding English texts in Common Voice~\cite{DBLP:conf/lrec/ArdilaBDKMHMSTW20} datasets to these speakers and synthesize the corresponding English-Chinese speech pairs.   We then set a threshold for WER and PESQ to filter the data.   Specifically, we remove data with a WER higher than 6\% or a PESQ below 3.9.   After filtering, we obtain a total of 277K English-Chinese speech pairs, comprising 263 and 124 hours of speech, respectively.

\vspace{-5pt}
\subsubsection{Unpaired Chinese TTS Data}
\vspace{-3pt}
To improve the generalization performance of our model and enable zero-shot capabilities, we employ a large amount of Chinese speech data with diverse acoustic conditions.   Specifically, we utilize an internal dataset along with two open-source datasets, AISHELL-3~\cite{DBLP:conf/interspeech/ShiBXZL21} and DiDiSpeech~\cite{DBLP:conf/icassp/GuoWJLZZLGZHL21}, which contain 500 hours of speech from 1,000 speakers of varying ages and genders, including professional speakers who record high-quality speech in a studio environment and ordinary speakers who record speech in typical rooms with ambient noise and slight reverberation.   Table~\ref{dataset_tts} provides details on the datasets.   From the total 500 hours of data, we randomly select 300 hours from 1K speakers for pre-training.


\begin{table}[h]
\centering
\vspace{-6pt}
\caption{Detail of the unpaired Chinese TTS datasets, containing the duration of all samples and number of speakers.}
\begin{tabular}{ccc}
\hline
Dataset             & Speakers & \begin{tabular}[c]{@{}c@{}}Hours\end{tabular} \\ \hline
Internal & 282       & 345                                                           \\
AISHELL-3~\cite{DBLP:conf/interspeech/ShiBXZL21}          & 218      & 85                                                            \\
DiDiSpeech~\cite{DBLP:conf/icassp/GuoWJLZZLGZHL21}          & 500      & 70                                                            \\ \hline
\end{tabular}
\vspace{-6pt}
\label{dataset_tts}
\end{table}

\vspace{-4pt}
\subsubsection{Test Sets}
\vspace{-3pt}
\begin{itemize}
\item \textbf{Seen speakers:} To evaluate the performance of our model, we randomly sample 200 utterances from the En2Zh parallel corpus to create separate validation and test sets.  We use the test set to assess the semantic translation and style transfer performance for the seen speakers.

\item \textbf{Unseen speakers:} 
Due to the limited number of bilingual speakers that can be synthesized by our TTS system, we utilize the English texts in the test set for seen speakers to synthesize speech spoken by English speakers with no overlap with the En2Zh parallel corpus.  Specifically, we generate a test set of 100 utterances from 20 speakers, containing expressions in different styles. 
\vspace{-3pt}
\end{itemize}

\vspace{-5pt}
\subsection{Model Configurations}
\vspace{-2pt}
\subsubsection{Evaluation Systems}
\vspace{-3pt}
Due to the lack of previous studies on zero-shot style transfer for direct S2ST, we mainly evaluate the performance of style transfer of our proposed methods compared to the original direct S2ST system. We conduct experiments on follows systems:

\begin{itemize}

\item \textbf{Direct S2UT: }Original direct S2UT system follows~\cite{DBLP:conf/interspeech/PopuriCWPAGHL22}. For consistency with our proposed model, we use acoustic model with vocoder structure instead of the unit decoder as the unit-to-speech module in this system.
\item \textbf{StyleS2ST: }Our proposed model.
\item \textbf{StyleS2ST-base: }Our proposed model without using unpaired Chinese data to pre-train the acoustic model. We use this system to validate the effectiveness of the pre-trained acoustic model.
\item \textbf{GT-StyleS2ST: }System using ground-truth discrete units in the unit-to-speech module of our proposed model.  We conduct this system to evaluate the topline of style transfer performance with no semantic error.
\vspace{-2pt}
\end{itemize}
\vspace{-5pt}
\subsubsection{Implementation Details}
\vspace{-3pt}

Our direct S2ST system is based on Fairseq~\cite{DBLP:conf/naacl/OttEBFGNGA19} and utilizes the Chinese HuBERT model\footnote{\url{https://huggingface.co/TencentGameMate/chinese-hubert-base}} for feature extraction. We employ the large-scale unsupervised Chinese speech dataset WenetSpeech~\cite{DBLP:conf/icassp/ZhangLGSYXXBCZW22} to train a k-means model with a vocabulary of 1000 units, using the 11-layer HuBERT feature. Subsequently, we encode all Chinese speech into discrete units using the k-means model. For the speech-to-unit module, we use the large wav2vec 2.0 speech encoder\footnote{\url{https://dl.fbaipublicfiles.com/fairseq/speech_to_speech/s2st_finetuning/w2v2/en/conformer_L.pt}} for English. Additionally, we utilize the discrete units extracted by the k-means model from the WenetSpeech dataset to train the unit BART decoder with the configuration in~\cite{DBLP:conf/interspeech/PopuriCWPAGHL22}.


In the unit-to-speech module, we follow the configurations in Fastspeech~\cite{DBLP:conf/nips/RenRTQZZL19} for acoustic model and DSPGAN-mm in DSPGAN~\cite{DBLP:journals/corr/abs-2211-01087} for vocoder. For the style adaptor, we adopt ECAPA-TDNN~\cite{DBLP:conf/interspeech/DesplanquesTD20} model configurations and train the model on speaker verification corpus, including the English dataset VoxCeleb 1\&2~\cite{DBLP:journals/corr/abs-2201-04583}, the Chinese dataset CN-Celeb 1\&2~\cite{DBLP:conf/icassp/FanKLLCCZZCW20}, and other English and Chinese datasets\footnote{\url{https://www.openslr.org/}}, with a total of more than 10K hours and 30K speakers.

\begin{table*}[tp]

\vspace{-4pt}
\centering
\caption{Experimental results of each system. MOS means naturalness MOS and SMOS means style similarity MOS. Higher is better for all indicators.}
\vspace{-5pt}
\resizebox{1\linewidth}{!}{ 
\begin{tabular}{lccccccccc}
\toprule
 &  \multicolumn{5}{|c|}{Seen Speakers}  & \multicolumn{4}{c}{Unseen Speakers} \\ \midrule
\multicolumn{1}{l|}{Systems}  & \multicolumn{1}{c|}{Units-BLEU} & \multicolumn{1}{c|}{ASR-BLEU}  & \multicolumn{1}{c|}{MOS} & \multicolumn{1}{c|}{SMOS} & \multicolumn{1}{c|}{SECS} & \multicolumn{1}{c|}{ASR-BLEU} & \multicolumn{1}{c|}{MOS} & \multicolumn{1}{c|}{SMOS} & \multicolumn{1}{c}{SECS}  \\ \midrule
\multicolumn{1}{l|}{Direct S2UT}  &\multicolumn{1}{c|}{58.21} &\multicolumn{1}{c|}{37.07} &\multicolumn{1}{c|}{3.66$\pm$0.06} &\multicolumn{1}{c|}{3.51$\pm$0.05} &\multicolumn{1}{c|}{0.774}  &\multicolumn{1}{c|}{18.21} &\multicolumn{1}{c|}{2.58$\pm$0.05} & \multicolumn{1}{c|}{2.15$\pm$0.06} & \multicolumn{1}{c}{0.646} \\
\multicolumn{1}{l|}{StyleS2ST}  &\multicolumn{1}{c|}{58.21} &\multicolumn{1}{c|}{\textbf{39.12}} &\multicolumn{1}{c|}{\textbf{3.76$\pm$0.05}} &\multicolumn{1}{c|}{3.83$\pm$0.06} &\multicolumn{1}{c|}{\textbf{0.844}} &\multicolumn{1}{c|}{30.15} &\multicolumn{1}{c|}{\textbf{3.43$\pm$0.06}} & \multicolumn{1}{c|}{\textbf{3.56$\pm$0.05}} & \multicolumn{1}{c}{\textbf{0.820}} \\
\multicolumn{1}{l|}{StyleS2ST-base}  &\multicolumn{1}{c|}{58.21} &\multicolumn{1}{c|}{38.85} &\multicolumn{1}{c|}{3.72$\pm$0.05} &\multicolumn{1}{c|}{\textbf{3.85$\pm$0.04}} &\multicolumn{1}{c|}{0.838}  &\multicolumn{1}{c|}{20.14} &\multicolumn{1}{c|}{2.57$\pm$0.05} & \multicolumn{1}{c|}{2.28$\pm$0.07} & \multicolumn{1}{c}{0.688}     \\ \midrule
\multicolumn{1}{l|}{GT-StyleS2ST}  &\multicolumn{1}{c|}{100.00} &\multicolumn{1}{c|}{58.42} &\multicolumn{1}{c|}{4.02$\pm$0.04} &\multicolumn{1}{c|}{3.89$\pm$0.03} &\multicolumn{1}{c|}{0.840}  &\multicolumn{1}{c|}{-} &\multicolumn{1}{c|}{-} & \multicolumn{1}{c|}{-} & \multicolumn{1}{c}{-}   \\ \midrule
\multicolumn{1}{l|}{Records}  &\multicolumn{1}{c|}{100.00} &\multicolumn{1}{c|}{65.50} &\multicolumn{1}{c|}{4.23$\pm$0.03} &\multicolumn{1}{c|}{4.09$\pm$0.03} &\multicolumn{1}{c|}{0.885}  &\multicolumn{1}{c|}{-} &\multicolumn{1}{c|}{-} & \multicolumn{1}{c|}{-} & \multicolumn{1}{c}{-} \\

\bottomrule

\end{tabular}
}
\label{MOS test}
\vspace{-8pt}
\end{table*}

\vspace{-5pt}
\subsection{Experimental Results}
\vspace{-3pt}
We evaluate each system for semantic translation and style transfer using various measures.   To assess the quality of semantic translation, we first compute the BLEU scores of translated discrete units, referred to as units-BLEU. Next, we use an internal Chinese ASR model to transcribe the synthesized speech to text and calculate the BLEU scores, which we refer to as ASR-BLEU. To evaluate the naturalness and style similarity of the synthesized speech, we conduct a mean opinion score (MOS) test across all the test sets for each system, with ten listeners rating each sample on a scale of 1 (worst) to 5 (best) for naturalness and style similarity.   Additionally, we use the speaker encoder cosine similarity (SECS) between the speaker embeddings extracted from the synthesized Chinese and the ground-truth English speech as an objective evaluation, with SECS scores ranging from 0 to 1.   A higher score indicates higher style similarity.

As shown in Table~\ref{MOS test}, our direct S2ST systems achieve good semantic translation accuracy.    Notably, ASR-BLEU is significantly lower than units-BLEU, and comparing the results of records and GT-StyleS2ST, we observe an adverse impact of the ASR system and synthetic speech quality to BLEU scores.    While the direct S2ST systems demonstrate generalization in semantic translation on unseen speakers, there is also a decrease in performance.    In the direct S2UT system and StyleS2ST-base system, the poor out-of-set generalization ability of the unit-to-speech module results in decreased intelligibility of synthesized speech, which is reflected in the lower ASR-BLEU score.    Regarding style transfer performance, we observe that the direct S2UT system achieves good style similarity within seen speakers, indicating that discrete units still contain some speaker-related information without the need for additional conditions.    Meanwhile, in other systems with a speaker adaptor, speaker similarity, and naturalness are significantly improved, indicating the ability of the speaker adaptor to represent speakers.    In the case of unseen speakers, both the direct S2UT and StyleS2ST-base models exhibit a significant decrease in naturalness and style similarity compared to the seen speakers, while our proposed StyleS2ST model maintains a better degree of naturalness and style similarity.    These results demonstrate the effectiveness of our pre-trained acoustic model and the direct S2UT and StyleS2ST-base models lacks the generalization performance for out-of-set timbre information.

\vspace{-5pt}
\subsection{Analysis on Acoustic Model Pre-training}
\vspace{-3pt}
To investigate the influence of data size and the speaker number used for pre-training the acoustic model in the zero-shot scenario, we conduct an analysis by varying the data size and speaker number and evaluating the naturalness and style similarity of the synthesized speech.
\vspace{-2pt}
\begin{table}[ht]
\centering
    \vspace{-3pt}
	\caption{Analysis on acoustic model pre-training in terms of naturalness \& style similarity.}
 \vspace{-3pt}
\resizebox{0.88\linewidth}{!}{ 
\begin{tabular}{l|c|c|c}
    \toprule
  & Naturalness  & \multicolumn{2}{c}{Style similarity}\\ \midrule
System & MOS & MOS & SECS \\ \midrule
  500hr+1000spk   & 3.40$\pm$0.04 & 3.47$\pm$0.05 & 0.815   \\
300hr+1000spk &  \textbf{3.43$\pm$0.05} & \textbf{3.56$\pm$0.06} &  0.820   \\
100hr+1000spk    & 2.96$\pm$0.05  & 3.41$\pm$0.06 & 0.808   \\
300hr+500spk    & 3.27$\pm$0.06  & 3.53$\pm$0.05 & \textbf{0.826}    \\
300hr+250spk    & 3.23$\pm$0.04  & 3.46$\pm$0.04 & 0.812    \\
StyleS2ST-base    & 2.57$\pm$0.06 & 2.28$\pm$0.07 & 0.688    \\
\bottomrule
\end{tabular}
}
\vspace{-6pt}
\label{pretraining_eval}
\renewcommand{\thefigure}{1}
\end{table}

 Results in Table~\ref{pretraining_eval} indicate that both factors have a significant influence on the performance of naturalness and style similarity of the model.
 It's interesting to observe that there brings a drop in performance when the data size increases from 300 to 500 hours with 1000 speakers. We deduce that this may result from model capacity limitations, which prevent the model from fully converging on the larger dataset.

\vspace{-4pt}
\section{Conclusions}
\vspace{-2pt}
In this paper, we propose StyleS2ST, aiming to achieve zero-shot style transfer in a direct S2ST system. We first build an English-to-Chinese direct S2ST system using a parallel corpus generated by a multi-lingual multi-speaker TTS model. Then we introduce a style adaptor to achieve style transfer and use a large amount of Chinese speech to pre-train the acoustic model to improve the zero-shot ability. Experimental results demonstrate that StyleS2ST can translate English speech to Chinese speech with good preservation of source speaking style to the target, even in zero-shot scenarios.  In future work, to further improve speech expressiveness in the translated speech, we would like to explore fine-grained style transfer~\cite{9693186}.

\clearpage

\bibliographystyle{IEEEtran}
\bibliography{mybib}

\end{document}